\documentclass[referee]{raa}            

\usepackage{graphicx,times}             
\usepackage{natbib}
\usepackage{makecell}
\usepackage{multirow}
\usepackage{amssymb,amsmath}
\usepackage{subcaption}
\bibpunct{(}{)}{;}{a}{}{,}

\usepackage{CJKutf8}

\usepackage[pagebackref=true]{hyperref}
\begin{document}
\begin{CJK*}{UTF8}{gbsn}

  \title{Digitization of Astronomical Photographic Plate of China and Astrometric Measurement of Single-exposure Plates}
  

   \volnopage{Vol.0 (20xx) No.0, 000--000}      
   \setcounter{page}{1}          

  \author
    {
    Zheng-Jun Shang (商正君)\inst{1}, Yong Yu (于涌)\inst{1,2}, Liang-Liang Wang (王亮亮)\inst{1}, Mei-Ting Yang (杨美婷)\inst{1}, Jing Yang (杨静)\inst{1}, Shi-Yin Shen (沈世银, 0000-0002-3073-5871) \inst{1,3}, Min Liu (刘敏)\inst{1}, Quan-Feng Xu (徐权峰)\inst{1,2}, Chen-Zhou Cui (崔辰州)\inst{2,4,5}, Dong-Wei Fan (樊东卫)\inst{2,4,5}, Zheng-Hong Tang (唐正宏)\inst{1,2}, Jian-Hai Zhao (赵建海)\inst{1}
    }  

   \institute{Shanghai Astronomical Observatory, Chinese Academy of Sciences, Shanghai 200030, China; {\it yuy@shao.ac.cn}\\
        \and 
            University of Chinese Academy of Sciences, Beijing 100049, China\\
        \and 
            Key Lab for Astrophysics, Shanghai, 200234, China\\            
        \and
            National Astronomical Observatories, Chinese Academy of Sciences, Beijing 100101, China\\
        \and
            National Astronomical Data Center, Chinese Academy of Sciences, Beijing 100101, China\\
    \vs\no
   {\small Received 20xx month day; accepted 20xx month day}}

\abstract{From the mid-19th century to the end of the 20th century, photographic plates served as the primary detectors for astronomical observations. Astronomical photographic observations in China began in 1901, and over a century, a total of approximately 30,000 astronomical photographic plates have been captured. These historical plates play an irreplaceable role in conducting long-term, time-domain astronomical research. To preserve and explore these valuable original astronomical observational data, Shanghai Astronomical Observatory has organized the transportation of plates taken at night from various stations across the country to the Sheshan Plate Archive for centralized preservation. For the first time, plate information statistics was performed. On this basis, the plates were cleaned and digitally scanned, and finally digitized images were acquired for 29,314 plates. In this study, using Gaia DR2 as the reference star catalog, astrometric processing has been carried out successfully on 15,696 single-exposure plates, including object extraction, stellar identification, and plate model computation. As a result, for long focal length telescopes, such as the 40cm double-tube refractor telescope and the 1.56m reflector telescope at the Shanghai Astronomical Observatory and the 1m reflector telescope at the Yunnan Astronomical Observatory, the astrometric accuracy obtained for their plates is approximately 0.1$^{\prime\prime}$ to 0.3$^{\prime\prime}$. The distribution of astrometric accuracy for medium and short focal length telescopes ranges from 0.3$^{\prime\prime}$ to 1.0$^{\prime\prime}$. The relevant data of this batch of plates, including digitized images and stellar catalog of the plates are archived and released by the National Astronomical Data Center. Users can access and download plate data based on keywords such as station, telescope, observation year, and observed celestial coordinates.
\keywords{astrometry---methods: data analysis---techniques: image processing---astronomical data bases: miscellaneous}
}
   \authorrunning{Z.-J. Shang, Y. Yu \& L.-L. Wang et al. }            
   \titlerunning{Digitization of Astronomical Plates and Astrometry of the single-exposure plates}  

    \maketitle
\section{Introduction}           
\label{sect:intro}

Before the widespread application of CCDs in the late 1980s, photographic plates played a crucial role as the primary astronomical imaging material. According to statistics, there are a total of 3 million astronomical photographic plates worldwide(\citealt{Grindlay+Griffin+2012}) which document celestial phenomena in observed regions that are no longer accessible. They are instrumental in conducting long-term, temporal-scale astronomical research. 

In 1901, a 40cm double-tube refractor telescope was installed in Sheshan, Shanghai, marking the beginning of astronomical photographic observations in China. Over a century, approximately 30,000 astronomical photographic plates were captured by observatories in China.

The photosensitive material of photographic plates is highly sensitive to the environment and gradually deteriorates over time. Only through digitization can astronomical plates be preserved for an extended period and conveniently utilized for scientific research. In 2000, the International Astronomical Union (IAU) established the "Plate Preservation and Digitization" working group. In 2018, the IAU General Assembly passed a resolution once again, urging global collaboration and joint efforts within the astronomical community to accelerate plate digitization. Currently, several research institutions in different countries have carried out astronomical plate digitization projects, including the DASH project at Harvard University Observatory(\citealt{Simcoe+etal+2006}), the APPLAUSE project jointly initiated by German observatories(\citealt{Tuvikene+etal+2014}), the NAROO project initiated by the Paris Observatory in France(\citealt{Robert+etal+2021}), and the D4A project at the Royal Observatory of Belgium(\citealt{DeCuyper+Winter+2005}).

From 2004, Chinese astronomers called for the digitization of astronomical plates (\cite{Jin+etal+2007}) and carried out preliminary preparatory work, including plate categorization, preservation, and preliminary analysis of scanning accuracy(\cite{Fu+Zhao+2010,Yu+etal+2013}). In 2012, China officially initiated the digitization of astronomical plates, with Shanghai Astronomical Observatory (SHAO) of the Chinese Academy of Sciences (CAS) organizing the transportation of astronomical photographic plates taken at night from various stations across the country to the Sheshan Plate Archive for unified preservation.

The organization of the information and the statistical work for the plates were also carried out. Between 2013 and 2017, SHAO collaborated with Nishimura Optical Manufacturing Co., Ltd. in Japan to develop a high-precision astronomical plate scanner with micrometer-level precision(\cite{Yu+etal+2017}). Using this equipment, the digitization of 29,314 astronomical plates in the inventory was completed. In recent years, based on the quality of the plates, the number of exposures, and the observed targets, astrometric processing of the first batch of 15,696 single-exposure plate images was carried out.

In this study, we give the first detailed introduction to the digitization of Chinese astronomical photographic plates. The focus of this study is to present a database for the China digitalized astronomical photographic plates, where the astrometric measurement processing of the 15,696 single-exposure plates has been carefully performed. The paper is organized as follows. Section 2 provides an overview of the Chinese astronomical plate data. Section 3 presents the digitization status of the plates. Section 4 describes the process and results of astrometric processing for digitized images of single-exposure astronomical plates. Section 5 introduces the basic information of the database access, and finally a conclusion is presented in Section 6.

\section{Overview of Chinese Astronomical Plate Data}
\label{sect:overiew}

The Sheshan Plate Archive houses 30,750 astronomical photographic plates captured by 11 telescopes from five observatories, namely SHAO, National Astronomical Observatories of the Chinese Academy of Sciences (NAOC), Purple Mountain Observatory (PMO), Yunnan Astronomical Observatory (YNAO), and Qingdao Observatory (QDO). These plates encompass a wide range of celestial objects, including the moon, asteroids, comets, binary stars, variable stars, eruptive stars, radio stars, star clusters, nebulae, and extragalactic objects (\cite{Jin+etal+2007}). Table~\ref{Tab1} presents the information on the telescope parameters, the span of observation years, plate sizes, and the number of plates. However, due to missing record information on some plates, the source of observation for 1,315 plates remains uncertain.

\begin{table}[b!]
\begin{center}
\caption[]{Information on Telescopes and Plates}\label{Tab1}
 \begin{tabular}{p{1.5cm}p{2.5cm}p{1.5cm}p{1cm}p{4.8cm}p{1cm}p{1cm}}  
  \hline
Observatory&Telescope&Focal Length(m)&Span&\makecell{Plate Size(cm)/\\Field View(degree)}&Telescope Code&Plate Quantity\\
  \hline
    \multirow{4}{*}{SHAO}&40cm double-tube refractor telescope&6.9&1901-1995&$18\times13$($1.5\times1.1$), $30\times24$($2.5\times2.0$)&01&6,741\\
    &1.56m reflector telescope&15.6&1986-1999&$20\times20$($0.7\times0.7$), $16\times16$($0.6\times0.6$), $15\times12$($0.6\times0.4$), $30\times24$($1.1\times0.9$)&02&263\\
 \hline
   \multirow{4}{*}{NAOC}&60/90cm Schmidt telescope&1.8&1964-1984&$16\times16$($5.1\times5.1$)&03&4,265\\
    &40cm double-tube refractor telescope&2.0&1964-1986&$30\times30$($8.5\times8.5$), $9\times12$($2.6\times3.4$), $16\times16$($4.6\times4.6$)&04&4,487\\
 \hline
    \multirow{4}{*}{YNAO}&1 meter reflector telescope&13.0&1979-1993&$16\times16$($0.7\times0.7$), $9\times6$($0.4\times0.3$)&05&1,017\\
     &60cm refractor telescope&7.5&1988&$9\times6$($0.7\times0.5$)&06&9\\
 \hline
    \multirow{5}{*}{QDO}&15cm refractor telescope(Xisha Yongxing)&0.75&1986-1987&$16\times16$($12\times12$)&07&79\\
    &32cm refractor telescope&3.68&1964-1997&$15\times12$($2.3\times1.9$), $16\times16$($2.5\times2.5$)&08&1,652\\
 \hline
    \multirow{6}{*}{PMO}&40cm double-tube refractor telescope&3.0&1964-1997&$30\times30$($5.7\times5.7$), $13\times18$($2.5\times3.4$), $12\times9$($2.3\times1.7$)&09&4,363\\
    &60cm refractor telescope&3.0&1954-1980&$12\times9$($2.3\times1.7$)&10&1,625\\
    &15cm refractor telescope&1.5&1949-1976&$9\times12$($3.4\times4.6$), $18\times13$($6.8\times5.0$), $10\times15$($3.8\times5.7$)&11&4,934\\
 \hline
    \multicolumn{6}{l}{The total quantity of astronomical plates*}&30,750\\
 \hline
 \end{tabular}
 *Due to missing information, the observation source of 1,315 astronomical photographic plates remains unknown.
\end{center}
\end{table}

\begin{figure}[h]
    \centering
    \includegraphics[width=0.5\textwidth, angle=0]{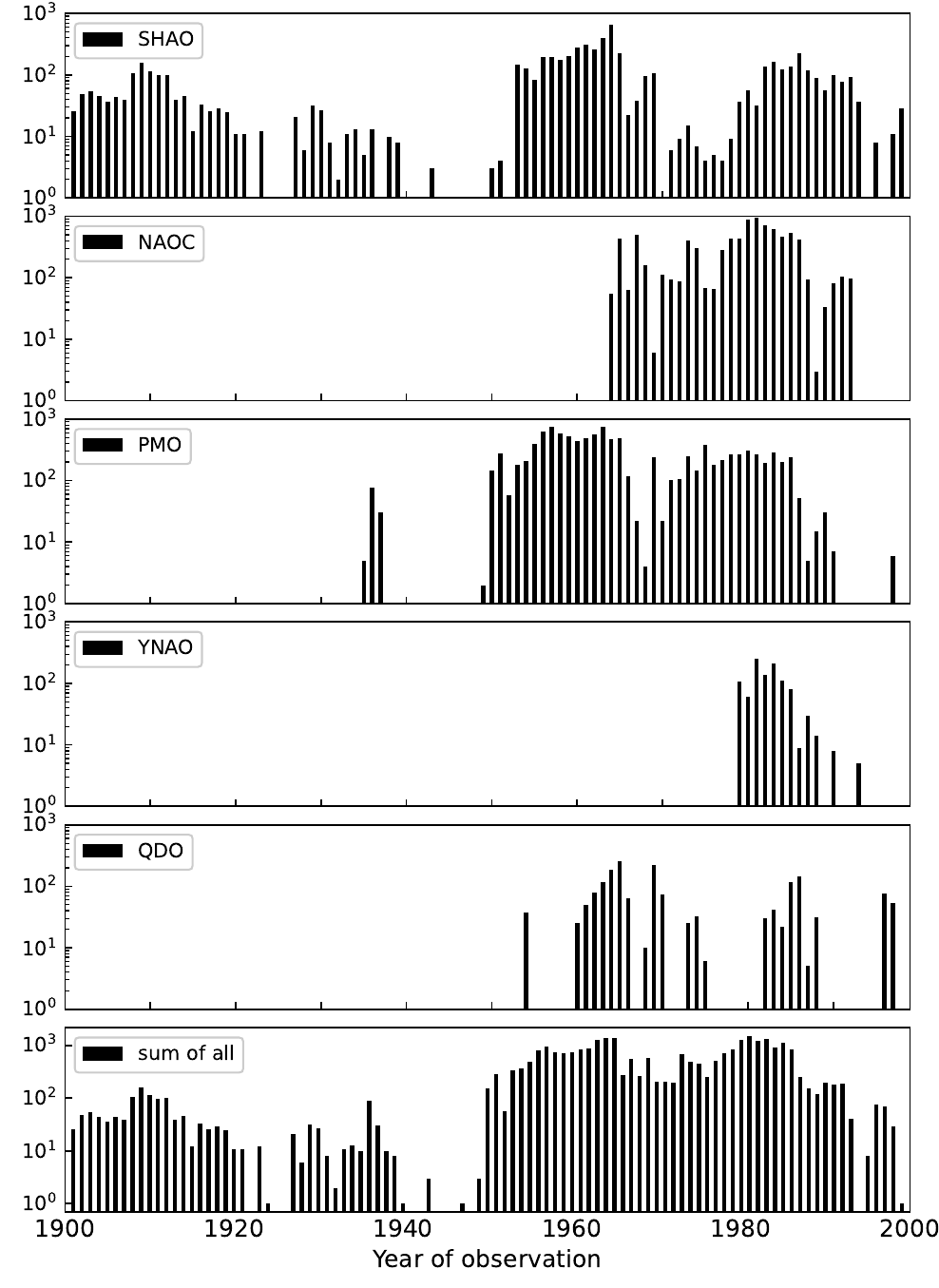}
    \caption{The distribution of the quantity of plates from each station over time. The top five panels depict the distribution of plate quantities for each station across different years, while the bottom panel represents the annual distribution of plates from all stations combined.}
    \label{Fig1}
\end{figure}

\begin{figure}[h]
    \centering
    \includegraphics[width=0.5\textwidth, angle=0]{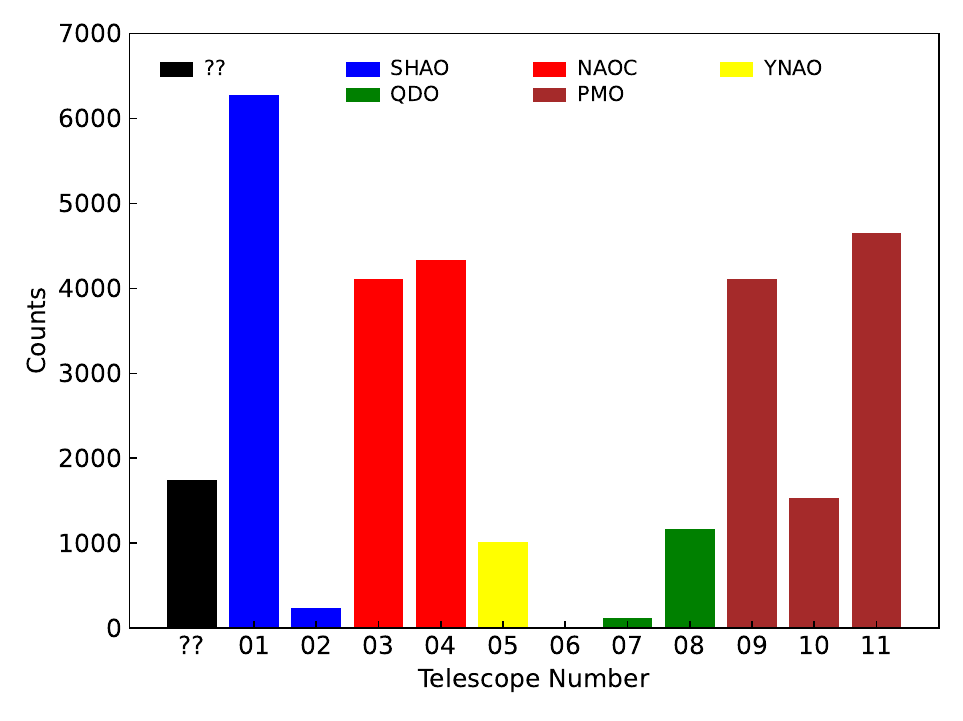}
    \caption{The distribution of the quantity of plates in the Sheshan Plate Archive by telescope. Different colors represent different observatories. The label ?? denotes plates with unidentified telescope information }
    \label{Fig2}
\end{figure}

\begin{figure}[h]
    \centering
    \includegraphics[width=0.5\textwidth, angle=0]{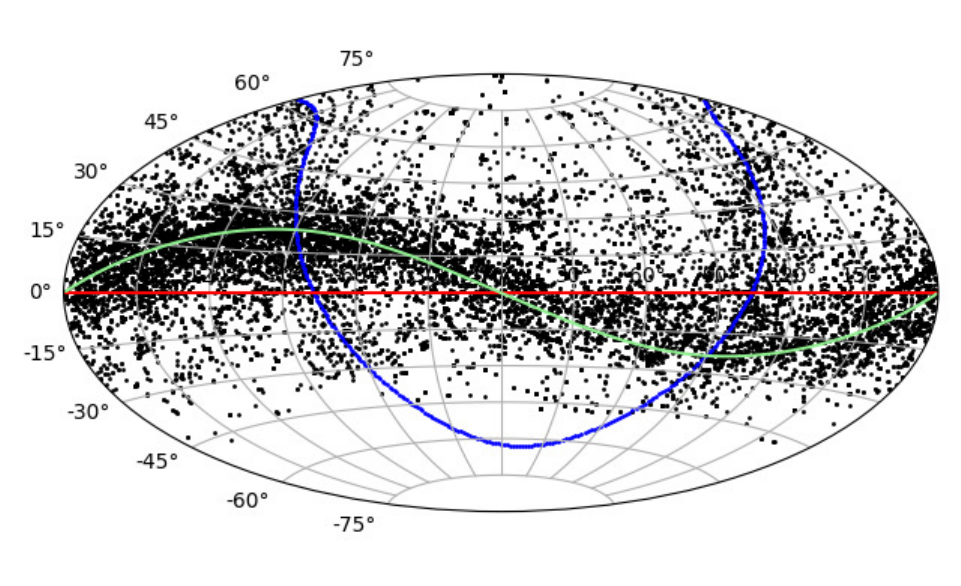} %
    \caption{Distribution of plate pointing for all plates. The red line represents the celestial equator, the green line represents the ecliptic, and the blue line represents the galactic plane}
    \label{Fig3}
\end{figure}

Figure~\ref{Fig1} shows the distribution of the quantities of astronomical plates from each station over time. It can be seen that SHAO had the longest duration of plate observation, spanning from 1901 to 1999. PMO had the highest number of observed plates, with a total of 10,922. Figure~\ref{Fig2} presents a histogram of the number of plates captured by different telescopes, with each color representing the corresponding observatory. Among these telescopes, the 40cm double tube refractor telescope (telescope No. 01), built in 1901, captured the highest number of plates, reaching 6,741. Figure~\ref{Fig3} illustrates the distribution of the pointings of all plates. It can be seen that a significant number of astronomical plates were taken along the celestial ecliptic, following the distribution of the zodiacal belt.

\begin{figure}[h]
    \centering
    \subfloat[]{
    \includegraphics[width=0.2\textwidth]{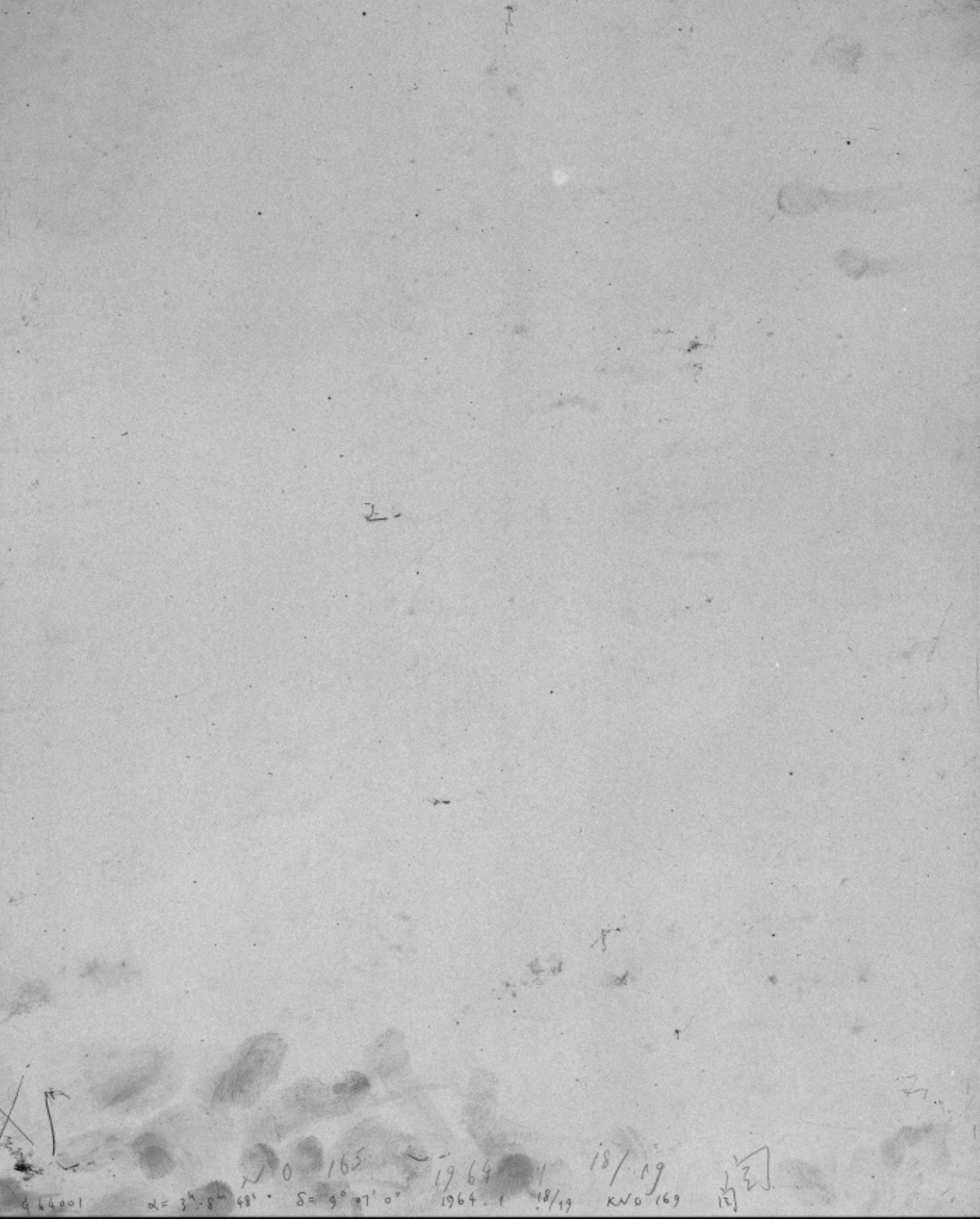}
    \label{Fig4(a)}
    }
    \subfloat[]{
    \includegraphics[width=0.2\textwidth]{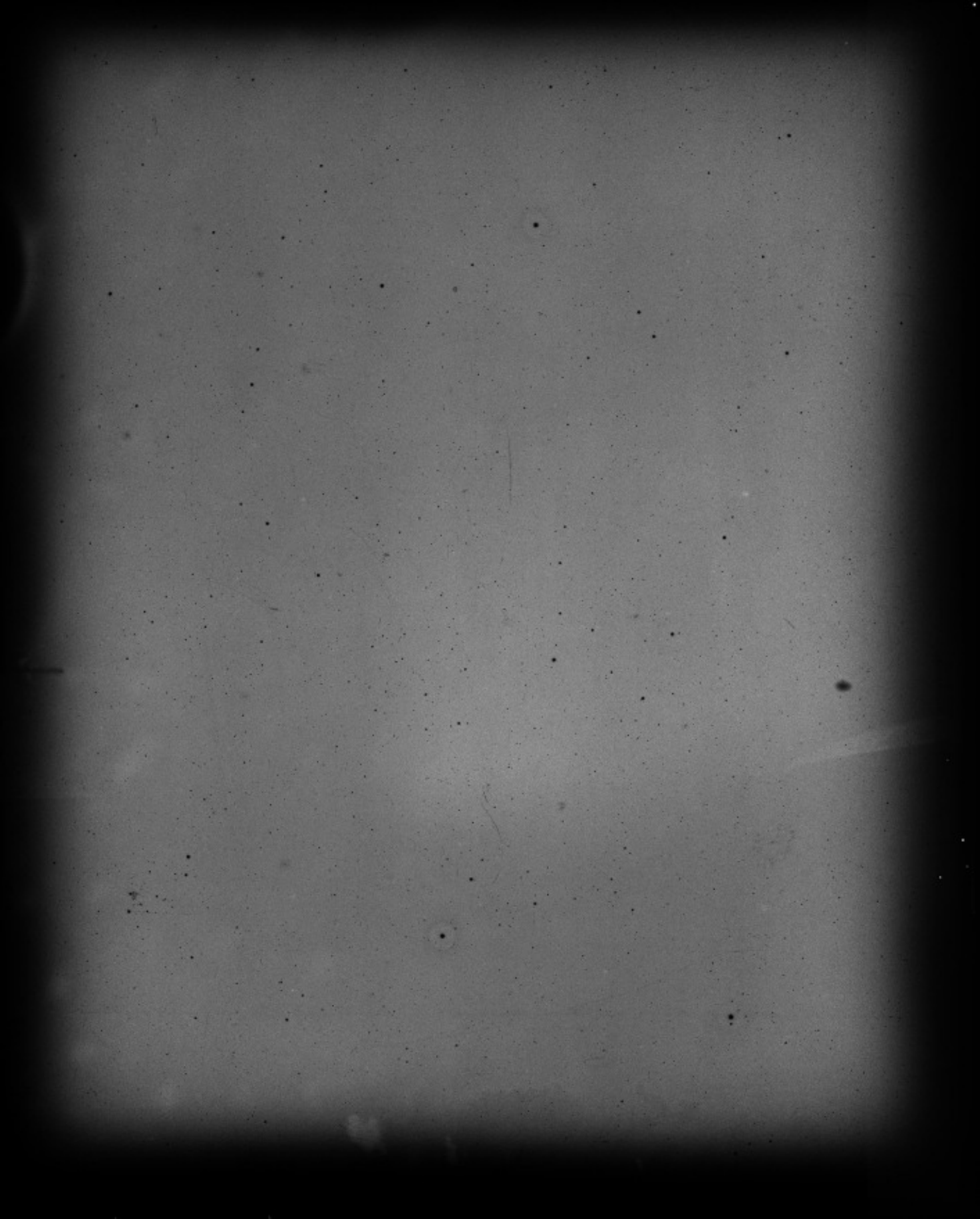}
    \label{Fig4(b)}
    }
    \subfloat[]{
    \includegraphics[width=0.2\textwidth]{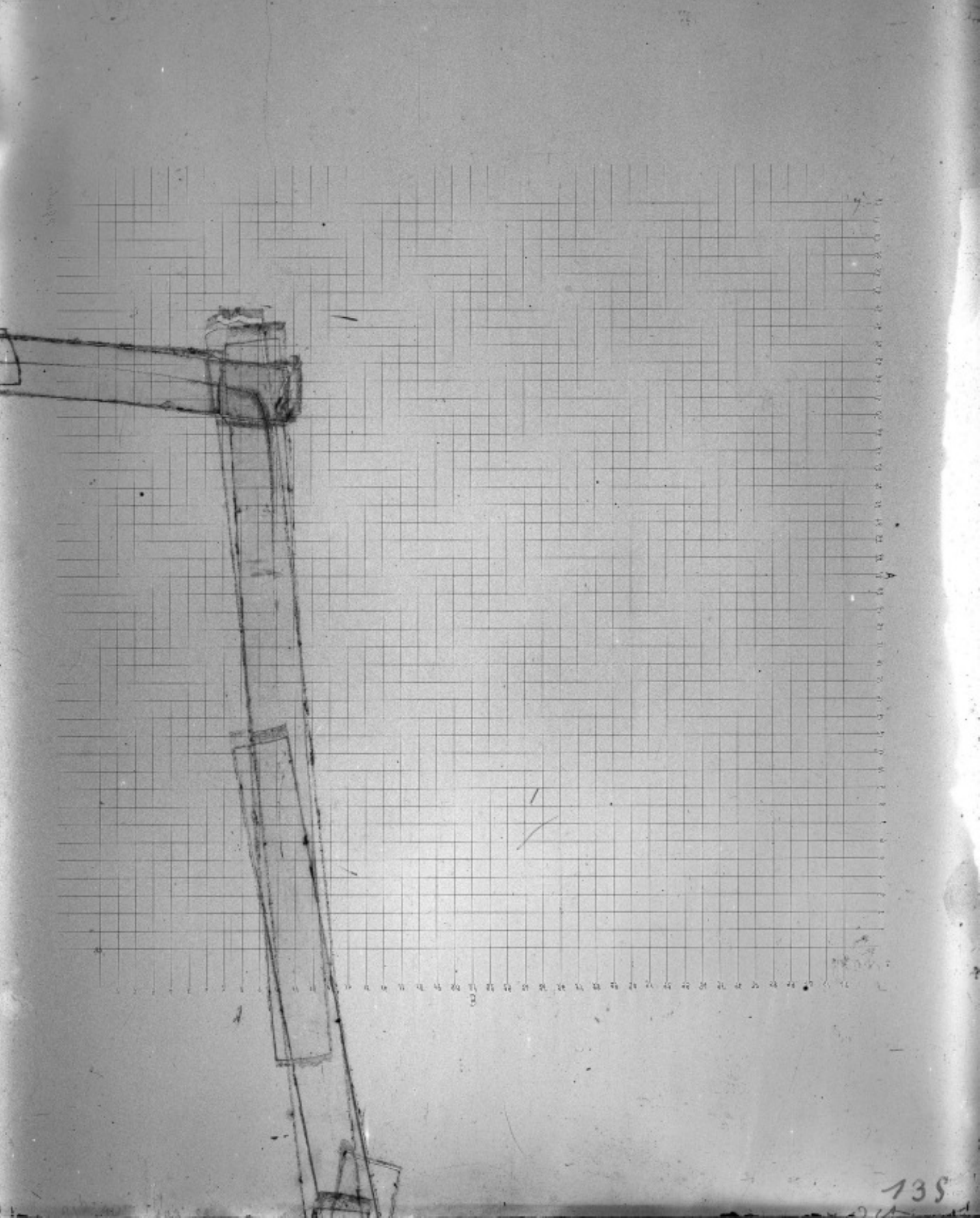}
    \label{Fig4(c)}
    }
    \subfloat[]{
    \includegraphics[width=0.2\textwidth]{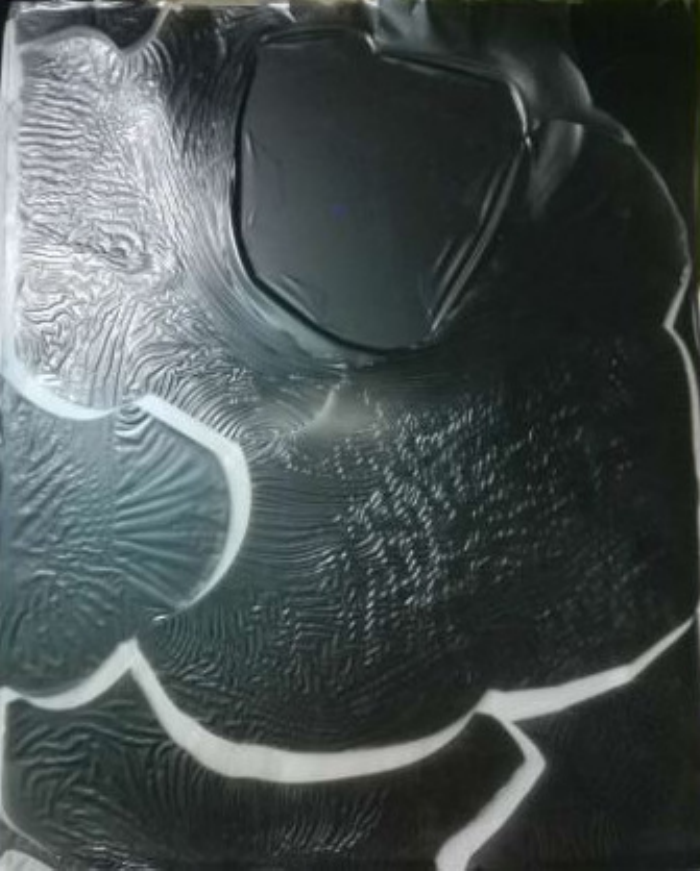}
    \label{Fig4(d)}
    }
    \caption{Examples of astronomical plates of different grades. (a)-(c): Scanned images of plates Grade 1 - Grade 3. (d): Actual photograph of a Grade 4 plate. }
    \label{Fig4}
\end{figure}

Due to their aging and various factors, there are significant differences in the quality of different plates. In June 2015, under the guidance of experts, the plates were roughly classified according to the condition of the glass and emulsion, the extent of the mold and the quality of the plate development. Plates with intact emulsion and no mold were classified as Grade 1. Plates with minor detachment, mold, or damage covering less than 25\% of the surface area were classified as Grade 2. Plates with detachment, mold, or damage covering less than 50\% of the surface area were classified as Grade 3. Plates with severe glass damage, extensive detachment of the emulsion, and significant mold were classified as Grade 4. Figure~\ref{Fig4} provides examples of plates of different grades. Table~\ref{Tab2} presents the number of plates in each grade. Grade 4 plates, due to severe damage, were not included in the subsequent digitization scanning process.
   
\begin{table}[h]
    \centering
    \caption[]{Quantity of Astronomical Plates for Different Quality Grades}
    \label{Tab2}
    \begin{tabular}{cc}
    \hline\noalign{\smallskip}
    Plate Quality Grade&Quantity of Plates\\
    \hline\noalign{\smallskip}
    1&11,475\\
    2&9,835\\
    3&8,004\\
    4&1,436\\
    \hline\noalign{\smallskip}
    \end{tabular}
\end{table} 

\section{Digitization of Astronomical Plates}
\label{sect:digitization}

To avoid the containment of dust particles, smudges, and other impurities in the scanned images, a cleaning process was performed on each plate prior to scanning. The emulsion side of the plate was cleaned using an air blower to remove surface dust. The glass side was wiped with a cotton cloth dipped in distilled water, followed by drying with a lint-free cloth(\cite{Yang+etal+2024}).

\begin{figure}[h]
    \centering
    \includegraphics[width=8cm]{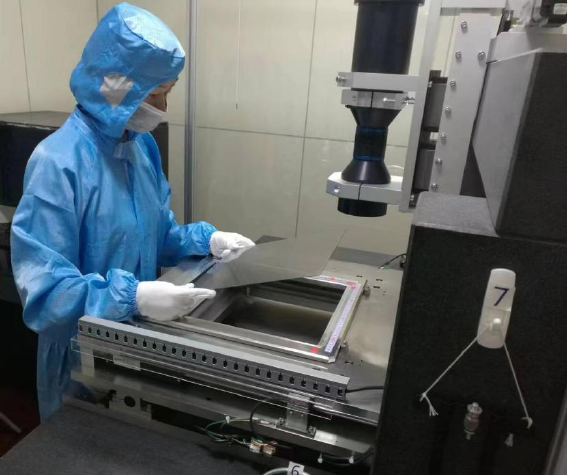}
    \caption{Staff performing astronomical plate scanning}
    \label{Fig5}
\end{figure}

Astronomical plate scanning was performed using a micrometer-level precision scanner (\citealt{Yu+etal+2017}) jointly developed by SHAO and Nishimura Optical Manufacturing Co., Ltd. in Japan, as shown in Figure~\ref{Fig5}. The scanner consists of a linear array camera, dual-sided telecentric lens, two-dimensional motion platform, and LED light source, all integrated on a granite platform. To mitigate the impact of environmental vibrations on plate scanning, the entire setup is placed on a separate base isolated from the surrounding floor. The scanning operation room is maintained at a constant temperature and humidity ($20^\circ C \pm 0.5^\circ C, 50\% \pm 5\%$) throughout the year. The scanner operates in a "line scanning" mode. Within a scanning range of $300mm \times 300mm$, the positional precision of the scans is better than 1 $\mu$m, and the scanning process takes no more than 10 minutes. The scanned images are saved as 16-bit FITS files.

To facilitate the query and retrieval of scanned astronomical plate images, a standardized naming convention was applied to each scanned image after the completion of the scanning process. Taking the name "SH8201S82072001" as an example, the naming convention for image files is listed in Table~\ref{Tab3}.

\begin{table}[h]
    \centering
    \caption[]{Naming Convention for Digitized Plate Images}
    \label{Tab3}
    \begin{tabular}{cc}
    \hline\noalign{\smallskip}
    String&Description\\
    \hline\noalign{\smallskip}
    SH&\makecell{Two uppercase letters indicate the observatory. SH for SHAO, \\BJ for NAOC, ZT for PMO, YN for YNAO, and QD for QDO.}\\ 
    \hline\noalign{\smallskip}
    82&\makecell{Two digits indicate the last two digits of the year when the \\plate was taken. If missing, @@ is used as a placeholder}\\
    \hline\noalign{\smallskip}
    01&\makecell{Two digits indicate the telescope No. If missing, \&\& is \\used as a placeholder}\\
    \hline\noalign{\smallskip}
    S82072&\makecell{Original plate No., if missing, \#\# is used as a placeholder}\\
    \hline\noalign{\smallskip}
    001&\makecell{Supplementary number, defaulting to 001. It is used to label \\multiple plates with identical preceding information}\\
    \hline\noalign{\smallskip}
    \end{tabular}
\end{table}

\section{Astrometry Processing of single-exposure plate images}
\label{sect:astrometry}

\begin{figure}[h]
    \centering
    \subfloat[]{
    \includegraphics[width=0.3\textwidth]{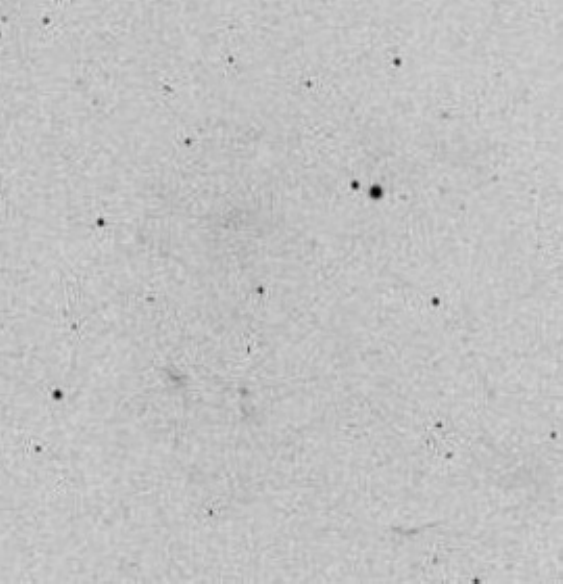}
    \label{Fig6a}
    }
    \subfloat[]{
    \includegraphics[width=0.3\textwidth]{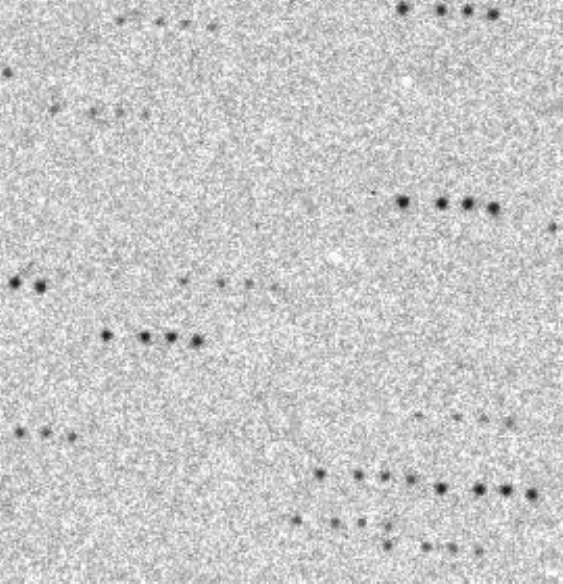}
    \label{Fig6b}
    }
    \subfloat[]{
    \includegraphics[width=0.3\textwidth]{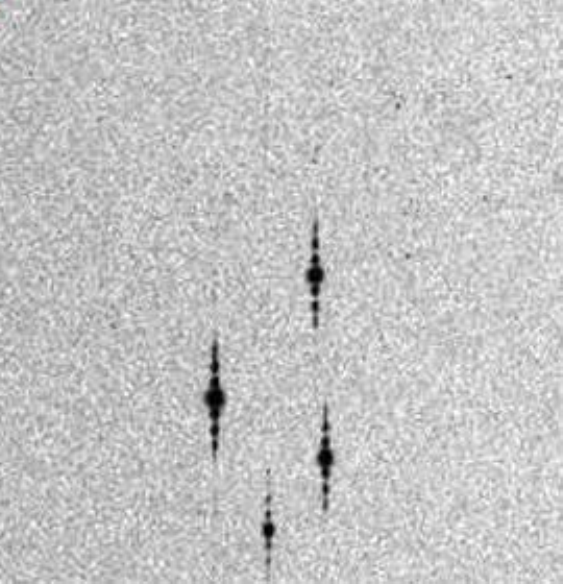}
    \label{Fig6c}
    }
    \\
    \subfloat[]{
    \includegraphics[width=0.3\textwidth]{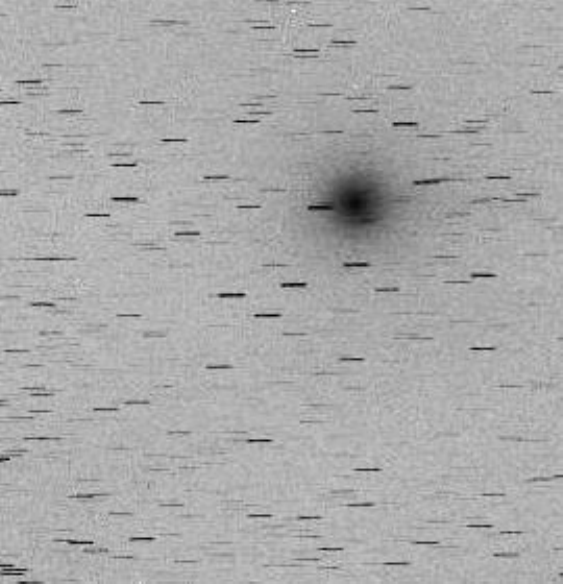}
    \label{Fig6d}
    }
    \subfloat[]{
    \includegraphics[width=0.3\textwidth]{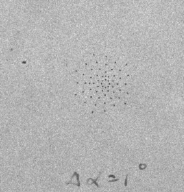}
    \label{Fig6e}
    }
    \caption{Examples of plates of different categories. (a) represents a single-exposure plate, (b) represents a multiple-exposure plate, (c) represents a grating observation plate, (d) represents a plate capturing near-Earth moving objects, and (e) represents a test observation plate.}
    \label{Fig6}
\end{figure}

Based on the imaging characteristics of stellar objects on plates, the plates were classified into the following categories: single-exposure plates (18,226), multiple-exposure plates (4,632), grating observation plates (364), near-Earth moving objects (such as major planets, minor planets, comets, etc.) plates (5,778), and test observation plates (314). Figure~\ref{Fig6} shows examples of plates from different categories, where Figure~\ref{Fig6a} corresponds to conventional single-exposure plates. Multiple-exposure plates refer to the repeated use of a single plate to capture multiple star images by slightly adjusting the telescope's pointing between each exposure, as shown in Figure~\ref{Fig6b}. Grating observation plates involves the use of diffraction gratings in the optical path of the telescope, resulting in multiple symmetrically distributed diffraction orders for each stellar object, as shown in Figure~\ref{Fig6c}. Near-Earth moving object plates are obtained by tracking moving targets with the telescope, where the target objects appear as circular shapes, while stellar objects exhibit elongation, as shown in Figure~\ref{Fig6d}. Test observation plates are captured to adjust telescope pointing, focal length, exposure time, and other parameters, as depicted in Figure ~\ref{Fig6e}.

In this paper, we present the astrometric processing on single-exposure plates.  Processing of other complex plates will be carried out in the follow-up work.

\begin{figure}[h]
    \centering
    \includegraphics[width=8cm, angle=0]{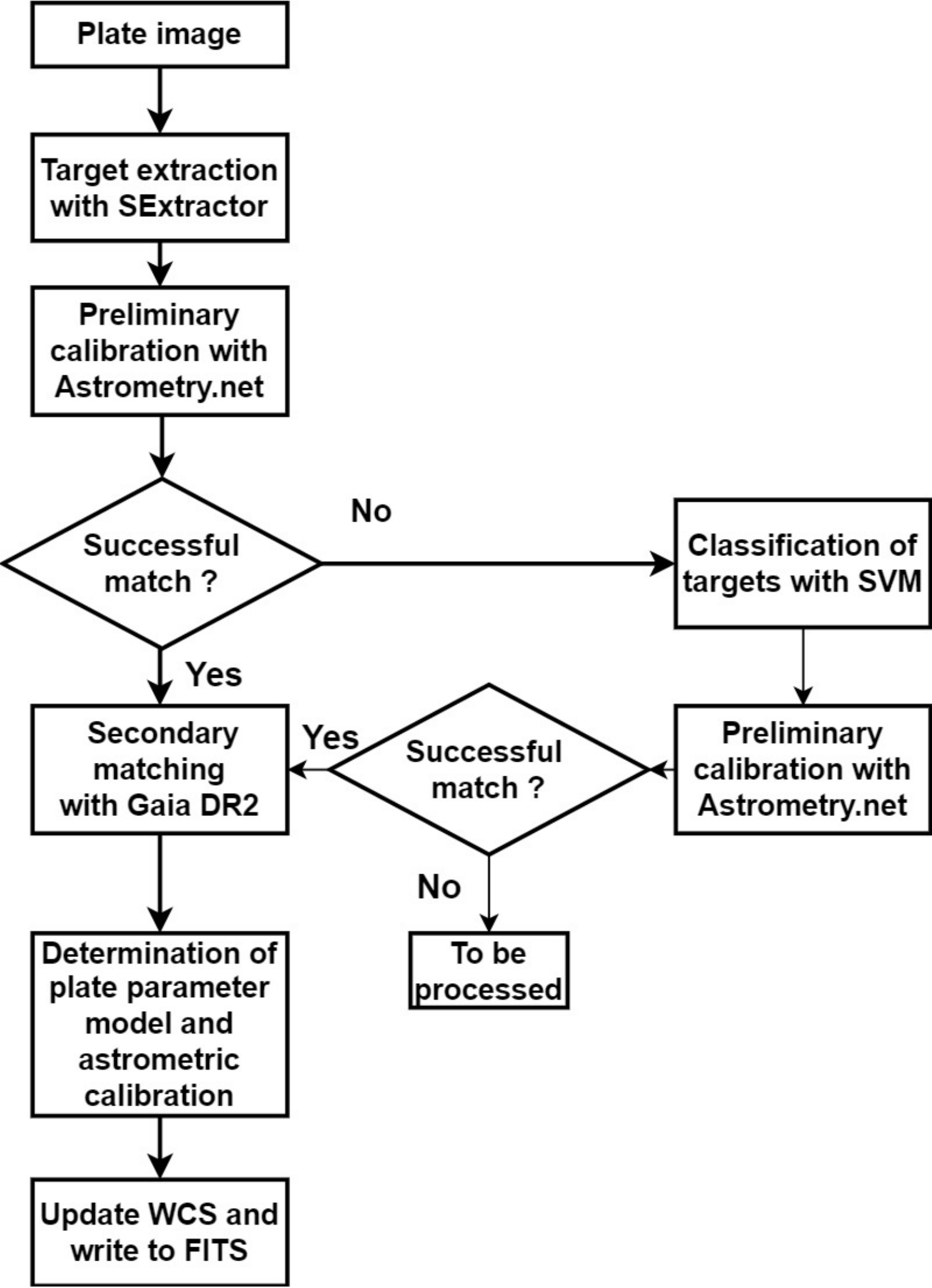} 
    \caption{Flow chart of astrometric calibtation for single-exposure astronomical plate images}
    \label{Fig7}
\end{figure}

Astrometric processing of plates involves several steps, including star extraction, reference star matching, plate parameter model computation, and celestial coordinate calculation. The workflow is illustrated in Figure~\ref{Fig7}. First, SExtracotr (\cite{Bertin+Arnouts+1996}) is used to obtain the measured coordinates and instrumental magnitudes of all candidate stellar objects on the plates. The parameter settings for the SExtractor software, tailored to the specific plates, are listed in Table~\ref{Tab4}. In the output parameters, we use XWIN\_IMAGE, YWIN\_IMAGE as measured coordinates of the objects, and MAG\_AUTO as the instrumental magnitude of the objects.

\begin{table}[h]
    \centering
    \caption{Config Values of Main Parameters in SExtractor}
    \label{Tab4}
    \begin{tabular}{cc}
    \hline\noalign{\smallskip}
    Parameter&Value\\
    \hline\noalign{\smallskip}
    DETECT\_THRESH&3.0\\
    DETECT\_MINAREA&5\\
    DETECT\_TYPE&CCD\\
    BACKPHOTO\_TYPE&LOCAL\\
    MASK\_TYPE&CORRECT\\
    CLEAN&Y\\
    FILTER&Y\\
    \hline\noalign{\smallskip}
    \end{tabular}
\end{table}

Then, the preliminary calibration of the detected objects is performed using Astrometry.net(\cite{Lang+etal+2010}). If the calibration is successful, the reference stars obtained from Astrometry.net are used to calculate the parameters of the plate model and determine the celestial coordinates of all stars. By matching with the Gaia DR2 catalog(\cite{GaiaCollaboration+2018, GaiaCollaboration+2016}), additional reference stars are identified.  This process is repeated until a stable match result with Gaia DR2 reference stars is achieved. For some plate images with many scratches, mold spots, and other artifacts, direct calibration using Astrometry.net may fail. In such cases, we employ the Support Vector Machine (SVM) method to filter out non-stellar objects from the extracted objects, which can improve the success rate of complex plate calibration for these challenging images. For more details, see Section 4.1.

The large physical size of astronomical plates (greater than $6cm \times 9cm$) makes their imaging susceptible to various factors such as optical distortions of the field of view, non-vertical alignment between the plate and telescope optical axis, poor atmospheric refraction, etc. Additionally, nonlinearity in response of plate to light intensity and asymmetric imaging resulting from telescope tracking errors or non-ideal imaging optical paths can lead to different center offsets for images of stars with different magnitudes, also known as magnitude equation(\cite{Girard+etal+1998}). These combined factors contribute to the imaging patterns of the plate. Therefore, through sampling and analysis of observational data from different telescopes, we have determined the appropriate calibration models for the respective telescope plate images. More details and analysis results can be found in Section 4.2. After combining the above methods, 15,696 single-exposure plates were astrometrically calibrated, and the calibration results are presented in Section 4.3.

To facilitate the query and retrieval of astronomical plate images, for each plate image, after completing the model computation for each plate, the model parameters are converted into World Coordinate System (WCS) parameters in the TAN-SIP format(\cite{Shupe+etal+2005}). These parameters include the celestial coordinates of the plate center, linear terms, and nonlinear terms, which are then written into the header of the image FITS file.

\subsection{Enhanced stellar target selection with SVM}

Due to the interference of scratches, mold spots, and impurities, a total of 4,714 single-exposure plate images cannot be used directly for initial reference star calibration using Astrometry.net. For these plates, we employ SVM for the classification of "stellar objects" and "none stellar objects". SVM is a machine learning method that aims to find an optimal hyperplane in a n-dimensional feature space by minimizing the objective function and effectively distinguishing between targets. SVM  offers advantages such as simplicity and strong generalization capabilities (\citealt{Noble+2006, Chauhan+etal+2019}), which has been widely used in the classification of celestial objects (\citealt{Zheng+etal+2005, Bu+etal+2014, Ma+etal+2016}).

\begin{figure}[h]
    \centering
    \includegraphics[width=8cm, angle=0]{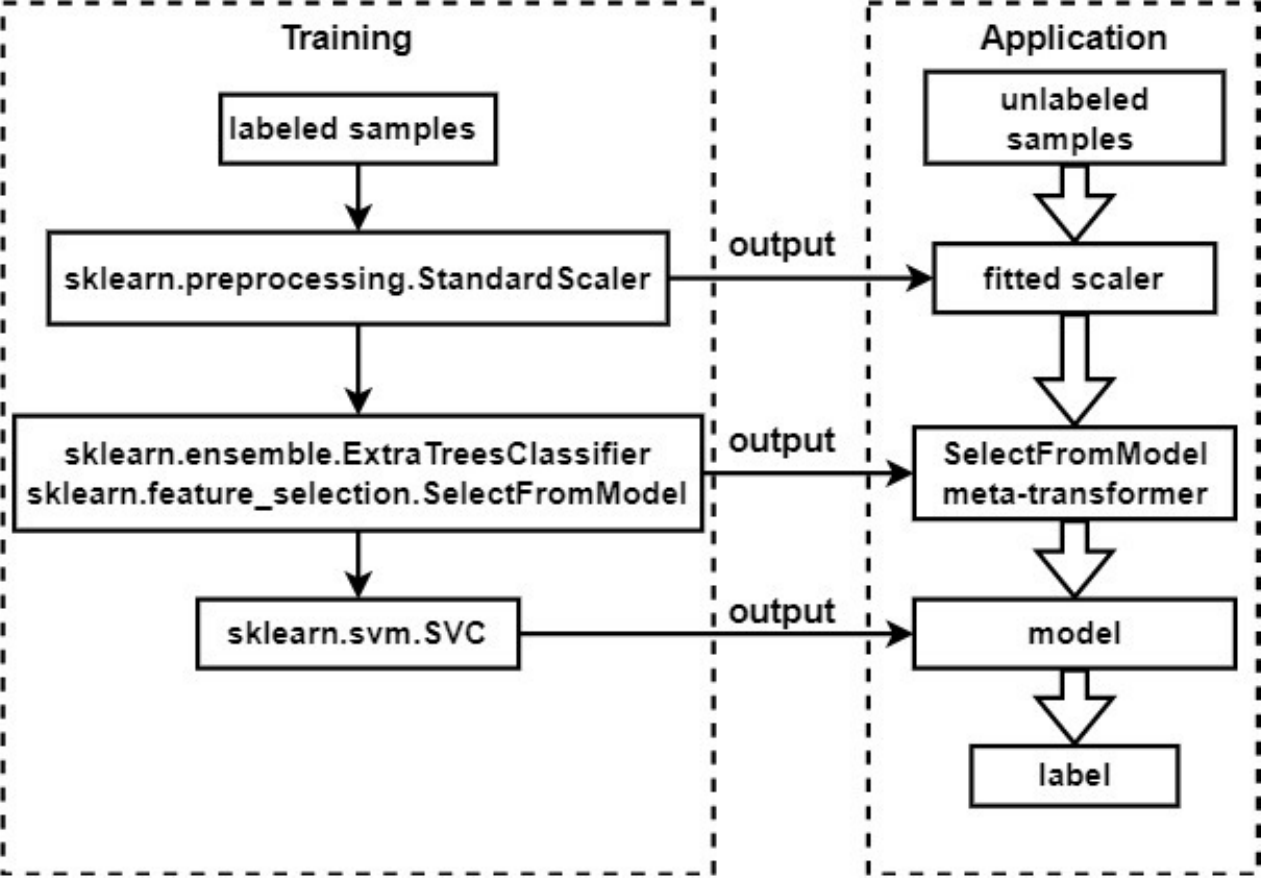}
    \caption{Training and application workflow of the SVM model}
    \label{Fig8}
\end{figure}

An SVM model was established using data from plates with completed astrometric calibration. SVM is implemented by calling scikit-learn (\cite{Pedregosa+etal+2011}), and the basic process is shown in Figure~\ref{Fig8}. In the first step, the parameters from the SExtractor output, excluding the constant and positional parameters, were selected as sample features, including the parameters related to MAG, AREA, ERROR and shape. Each extracted object consists of a data sample in the SVM algorithm. In the second step, the samples are standardized to ensure that each feature has a similar impact on the model's classification ability, avoiding individual features dominating the model's classification ability. In the third step, feature selection is performed by constructing an extremely randomized forest classifier to rank the features based on their importance. Features that exceed a threshold are selected for subsequent training and application. The selected feature parameters and sample labels are then used for model training. Finally, the model is applied to plates where the initial calibration using Astrometry.net has failed, and the extracted objects are labeled as "stellar objects" or "non-stellar objects".

\begin{figure}
    \centering
    \includegraphics[width=0.5\textwidth]{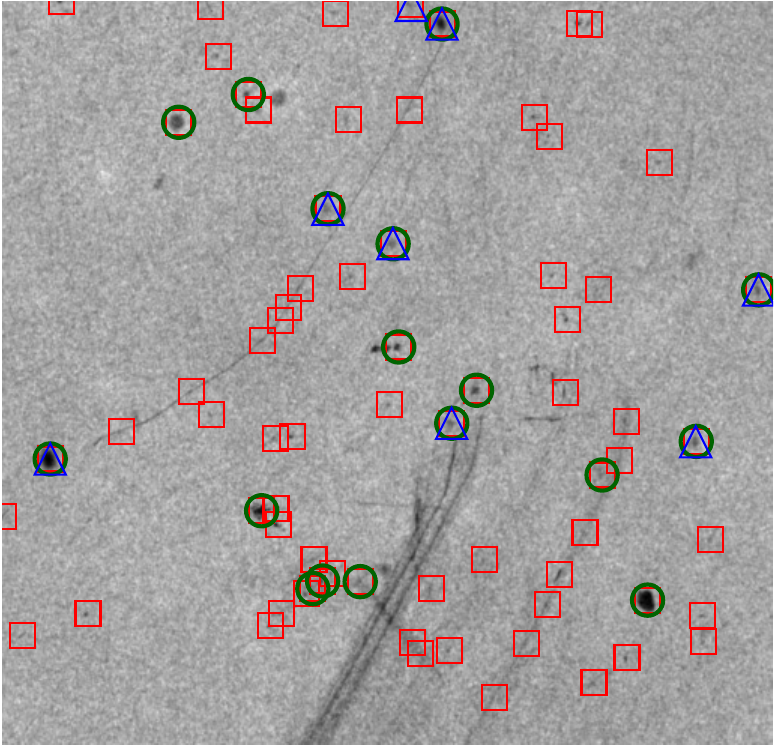}
    \caption{Illustration of target classification for plate SH6201V62036001. The red squares represent the targets extracted by the SExtractor, the green circles represent the targets classified as stellar objects by the SVM, and the blue triangles represent the targets matched with the Gaia DR2.}
    \label{Fig9}
\end{figure}

Randomly taking 10 plates as an example, Table~\ref{Tab5} shows the proportion of true stellar objects before and after SVM classification. $n_2/n_1$ represents the proportion of true stellar objects among objects directly extracted from SExtractor. $n_4/n_3$ represents the proportion of true stellar objects among targets classified as "stellar objects" by SVM. It can be observed that the proportion of true stellar objects significantly increases after the SVM classification. Using the plate SH6201V62036001 as an example, Figure~\ref{Fig9} displays the SExtractor extraction results,  the SVM classification results, and the matching results with the Gaia DR2 star catalog.

\begin{table}[h]
    \begin{center}
    \caption{Proportion of Stellar Objects in the Targets Before and After SVM Labeling(using 10 plate images as an example)}\label{Tab5}
    \begin{tabular}{ccccccc}
    \hline\noalign{\smallskip}
    Plate Name      &$n_1$    & $n_2$   & $n_2/n_1$& $n_3$   &  $n_4$ & $n_4/n_3$\\
    \hline\noalign{\smallskip}
    SH6201V62027001 &33,957 & 4,400 & 0.13 & 5,607 & 2,931& 0.52\\
    SH6201V62036001 &57,650 & 6,302 & 0.11 & 11,902& 4,979& 0.42\\
    ZT7509III267001 &57,30  & 716   & 0.12 & 643   & 336  & 0.52\\
    SH6201V62021001 &57,375 & 5,039 & 0.09 & 6,679 & 3,718& 0.56\\
    ZT6111NO65001   &2,294  & 176   & 0.08 & 234   & 100  &0.43\\
    SH5901G59041001 &9,986  &1,191  & 0.12 & 2,526 & 1,059&0.42\\
    SH5701CL57039001&21,520 &1,948  & 0.09 & 3,907 & 1,617&0.41 \\
    SH0501145001    &62,407 &8,181  & 0.13 & 14,779& 6,668&0.45\\
    SH8201CL82006001&88,887 &4,971  & 0.06 & 5,915 & 3,681&0.62\\
    BJ8303SD1867001 &238,487&11,166 & 0.05 & 10,067& 9,487&0.94\\
    \hline
    \end{tabular}\\
    \end{center}
    *$n_1$ is the number of targets extracted by the SExtractor. $n_2$ is the number of targets matched to Gaia DR2. $n_3$ is the number of targets labeled as stellar objects by SVM. $n_4$ is the number of true stellar objects (matched to Gaia DR2) among the targets labeled as stellar objects by SVM.
\end{table}
    
After SVM target selection, Astrometry.net was used again for initial calibration, and 2,184 calibrated results were successfully obtained from 4,714 plate images. Therefore, we finally have 15,696 single-exposure plates (18226 - 4714 + 2184) passed this initial step of astrometric processing, which constitutes the astronomical photographic plate database presented in this study.

\subsection{Selection of Plate Parameter Model}

\begin{figure}[h]
    \centering
    \includegraphics[width=0.5\textwidth]{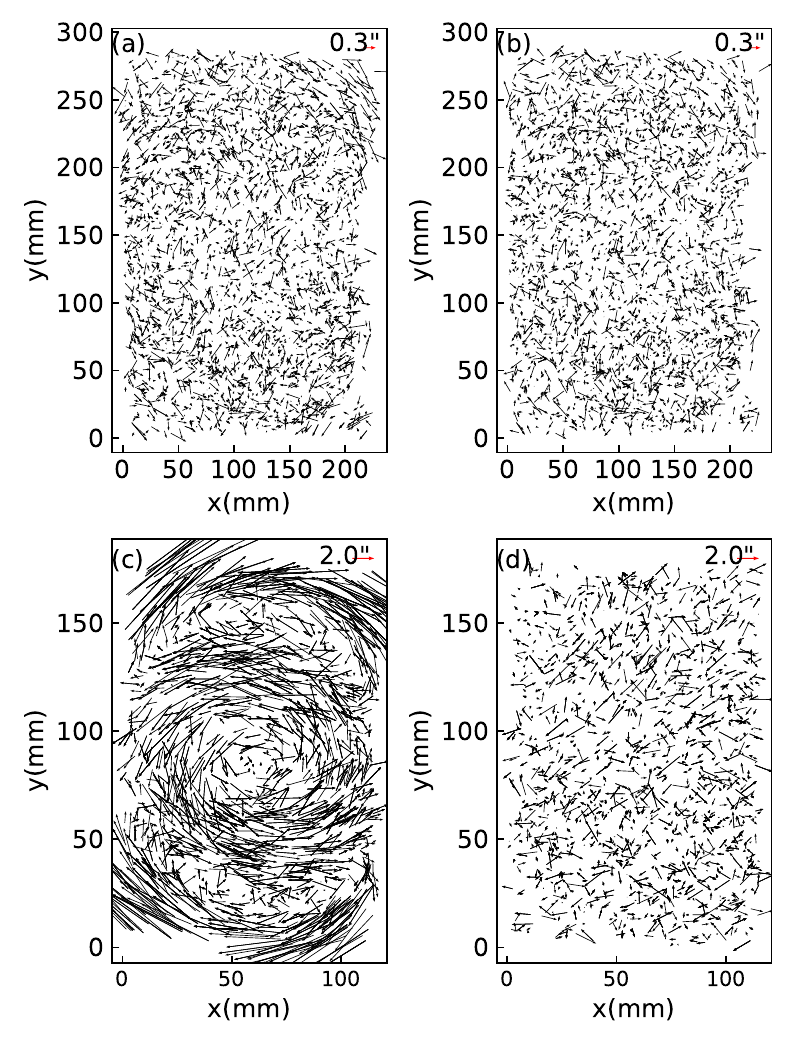}
    \caption{Distribution of reference star residuals on the image. (a) and (b) represent the case of plate SH6301V63096001 taken with telescope No. 01 at SHAO, using 1st- and 2nd-order model reduction, respectively. (c) and (d) represent the case of plate ZT5011T123001 taken with telescope No. 11 at PMO, using 2nd and 3rd-order model reduction respectively.}
    \label{Fig10}
\end{figure}

The plate parameter model is used to describe the relation between the ideal coordinates and measured coordinates of stellar objects on the plates, including translation, rotation, scale, and distortion. To determine the appropriate form of the plate parameter model, a random sample of 100 plates from each telescope was taken. The plates were individually reduced using conventional 1st, 2nd, 3rd, and 4th-order models (i.e., 6, 12, 20, and 30-parameter models, respectively), and the standard deviation of the residuals concerning the Gaia DR2 reference stars were examined for each plate model reduction. The best model is chosen based on the criteria of having the lowest order, lowest standard deviation, and randomly distributed residuals. Table~\ref{Tab6} presents the standard deviation of the reference star residuals for each telescope using different models. It can be seen that for the plates taken with telescope No. 01 (40cm double refractor telescope at SHAO), the standard deviation of the 2nd-order model is comparable to that of higher-order models. For plates taken with other telescopes, the standard deviation of the 3rd-order model is comparable to that of higher-order models. Taking plate SH6301V63096001 from telescope No. 01 and plate ZT5011T123001 from telescope No. 11 as examples. Figure~\ref{Fig10} shows the distribution of the reference star residuals in the image. It can be seen that the residual distribution is predominantly random. Therefore, except for the plates taken with telescope No. 01 at SHAO, which are reduced using a 2nd-order plate parameter model, plates from other telescopes are reduced using the 3rd-order plate parameter model.
    
\begin{figure}[h]
    \centering
    \includegraphics[width=0.4\textwidth]{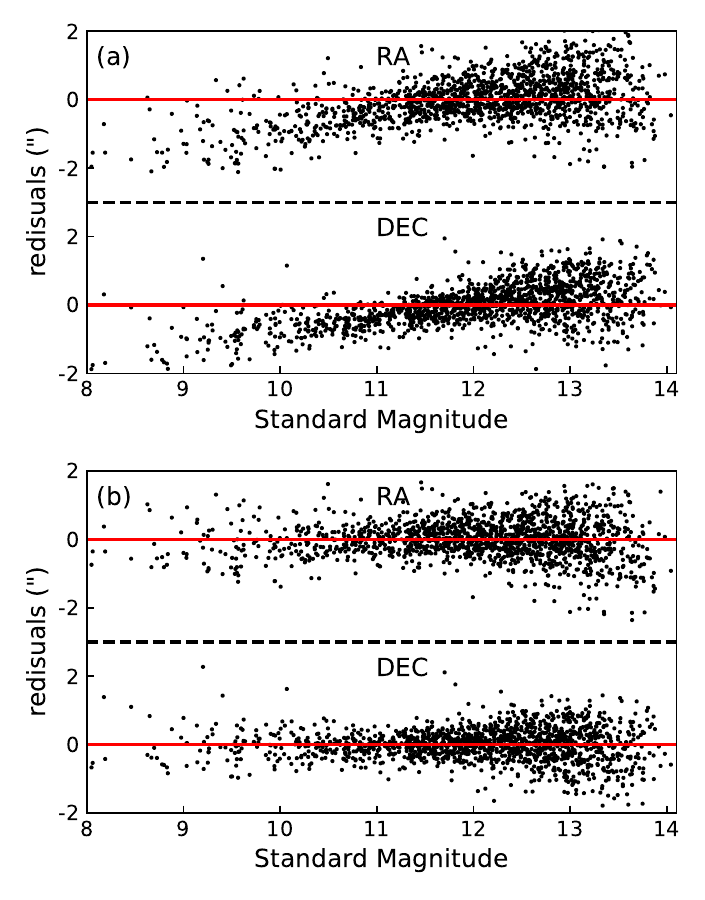}
    \caption{Distribution of residuals of reference stars with magnitude. Data from plate ZT5011T123001 taken with telescope No. 11 at PMO, with the reduction of a 3rd-order model. The upper panels are residuals of declination and the lower are of right ascension. (a) shows the distribution of residuals without magnitude equation correction. (b) shows the distribution  of residuals with magnitude equation correction.}
    \label{Fig11}
\end{figure}

A direct consequence of the magnitude equation is that the position offsets are related to the stellar magnitudes. Taking the plate ZT5011T123001 as an example, Figure~\ref{Fig11}(a) illustrates the distribution of the residuals of reference stars as a function of their magnitudes. Note that the magnitude used here is the $g$  magnitudes of the reference stars in the Gaia DR2 catalog. It can be seen that the distribution exhibits significant systematic patterns, with a maximum value of up to 2\arcsec. To mitigate the impact of this factor, we introduced a magnitude correction term in the reduction of the plate parameter model, which is $p_{1}\cdot mag + p_{2}\cdot mag^{2}$, where $mag$ represents the magnitude of the reference star, and $p_{1}$ and $p_{2}$ are the coefficients to be determined. Figure~\ref{Fig11}(b) displays the distribution of the residuals of the reference stars after incorporating the magnitude correction term. The correlation between the residuals and stellar magnitudes is significantly reduced.

\begin{table}[t]
    \centering
    \caption{Standard Deviations of Reference Star Residuals Under Different Order Plate Parameter Models}
    \label{Tab6}
    \begin{tabular}{cccccc}
    \hline\noalign{\smallskip}
    \multirow{3}{*}{Telescope No.}&\multicolumn{4}{c}{Standard Deviation(")}&\multirow{3}{*}{\makecell{Mean Quantity of\\ Reference Stars}}\\
    \cline{2-5}
    ~&\makecell{1st-order\\ model}&\makecell{2nd-order\\ model}&\makecell{3rd-order\\ model}&\makecell{4th-order\\ model}&~\\
    \cline{2-5}
    01& 0.138&  0.125&  0.123&  0.122& 1,996\\
    02& 0.149&  0.132&  0.097&  0.097& 768\\
    03& 0.525&	0.439&  0.349&	0.345& 27,818\\
    04&	3.386& 	2.098& 	0.593& 	0.592& 	65,228\\
    05& 0.220& 	0.213& 	0.107& 	0.105& 	454\\
    08&	0.383& 	0.334& 	0.308& 	0.306& 	2,184\\
    09&	0.718& 	0.600& 	0.271& 	0.271& 	21,116\\
    10&	0.761& 	0.721& 	0.593& 	0.587& 	1,559\\
    11&	2.790& 	2.265& 	0.599& 	0.596& 	3,169\\
    \hline\noalign{\smallskip}
    \end{tabular}
\end{table}
    
\subsection{Astrometric Calibration Results}

\begin{figure}[h]
    \centering 
    \includegraphics[width=0.7\textwidth]{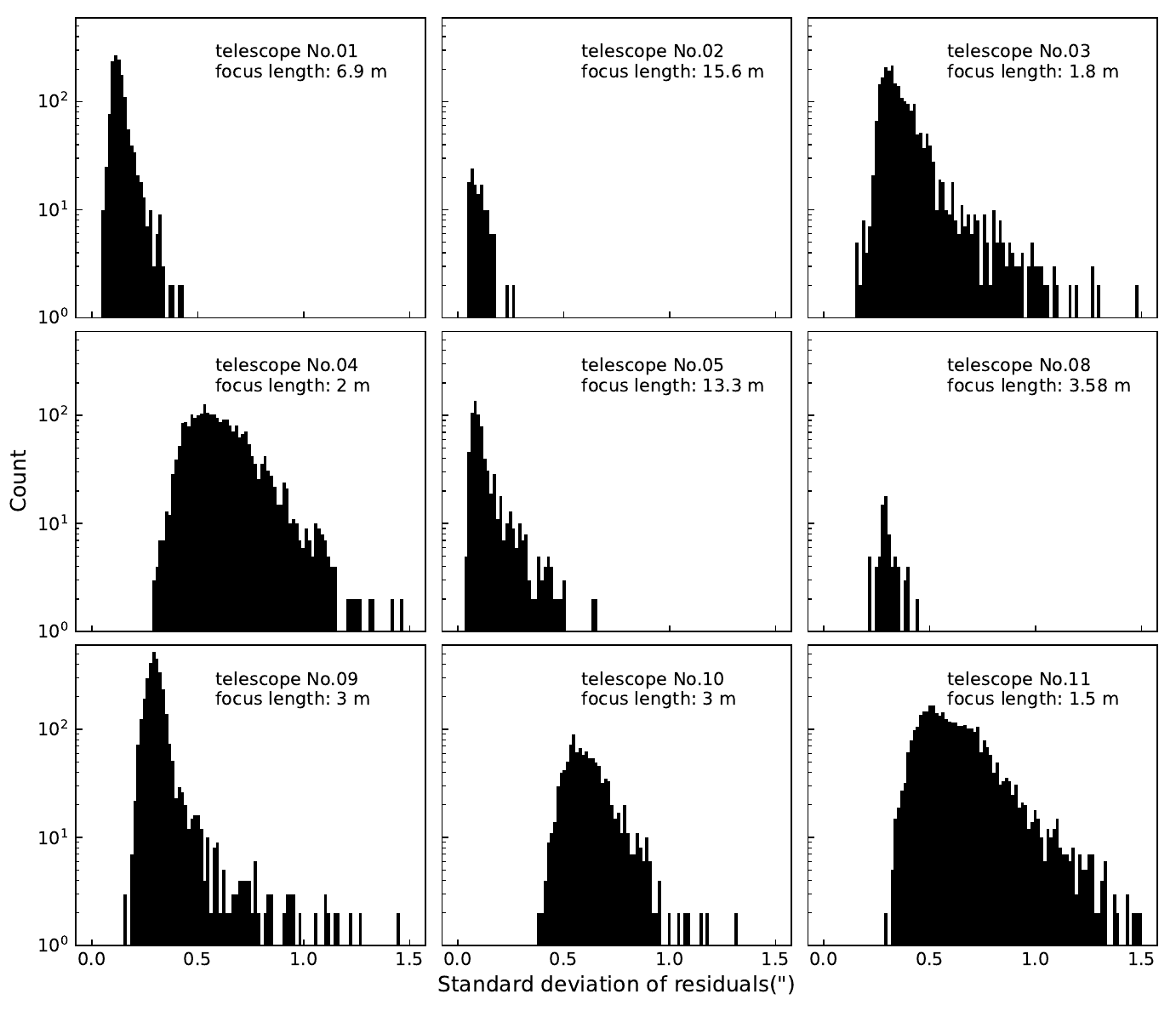}
    \caption{Distribution of standard deviation of the positional residuals of reference stars on each plate by telescope. None of plates from telescope No. 06 and 07 were astrometric calibrated successfully, so there is no data from telescope No. 06 and 07. }
    \label{Fig12}
\end{figure}

Using Gaia DR2 as the reference star catalog, astrometric calibration was conducted on a total of 15,696 single-exposure plates, obtaining the celestial coordinates of the stellar objects on each plate in the International Celestial Reference System (ICRS). Figure~\ref{Fig12} presents the distribution of standard deviation of the residuals of the reference stars on each plate by telescope. The standard deviation reflects the astrometric accuracy of the stars on the plates, mainly influenced by the focal length of the telescope and possible by the condition of plate preservation. For telescopes with longer focal lengths, such as the 1.56m reflecting telescope(No. 01) and the 40cm double tube refracting telescope (No. 02) at SHAO, and the 1m reflecting telescope(No. 05) at YNAO, the accuracy distribution of the plates ranges from 0.1\arcsec to 0.3\arcsec. For telescopes such as the 60/90cm Schmidt telescope(No. 03) at NAOC, the 32cm refracting telescope(No. 08) at QDO, and the 40cm double-tube refracting telescope(No. 09) at PMO, the distribution of the accuracy of the plates ranges from 0.3\arcsec to 0.5\arcsec. Finally, for the 40cm double tube refracting telescope (No. 04) at NAOC, the 60 cm reflecting telescope (No. 10) and the 15 cm reflecting telescope (No. 11) at PMO, the distribution of the accuracy of the plates ranges from 0.4\arcsec to 1.0\arcsec.

\begin{figure}[t]
        \centering
        \includegraphics[width=0.5\textwidth]{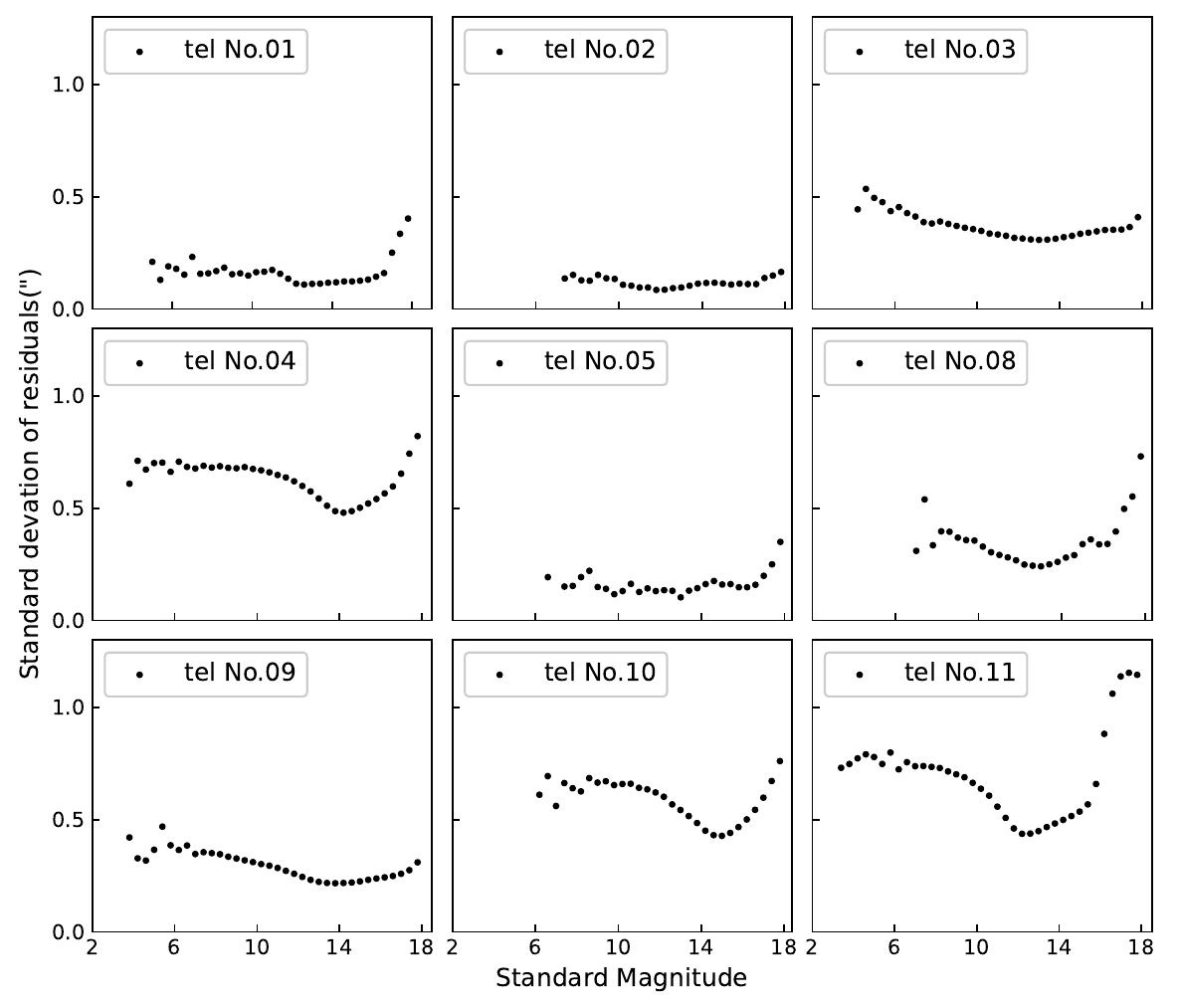}
        \caption{Distribution of the standard deviation of the positional residuals of the reference stars by telescope with magnitude.}
        \label{Fig13}
    \end{figure}

\begin{figure}[h]
        \centering
        \includegraphics[width=0.5\textwidth]{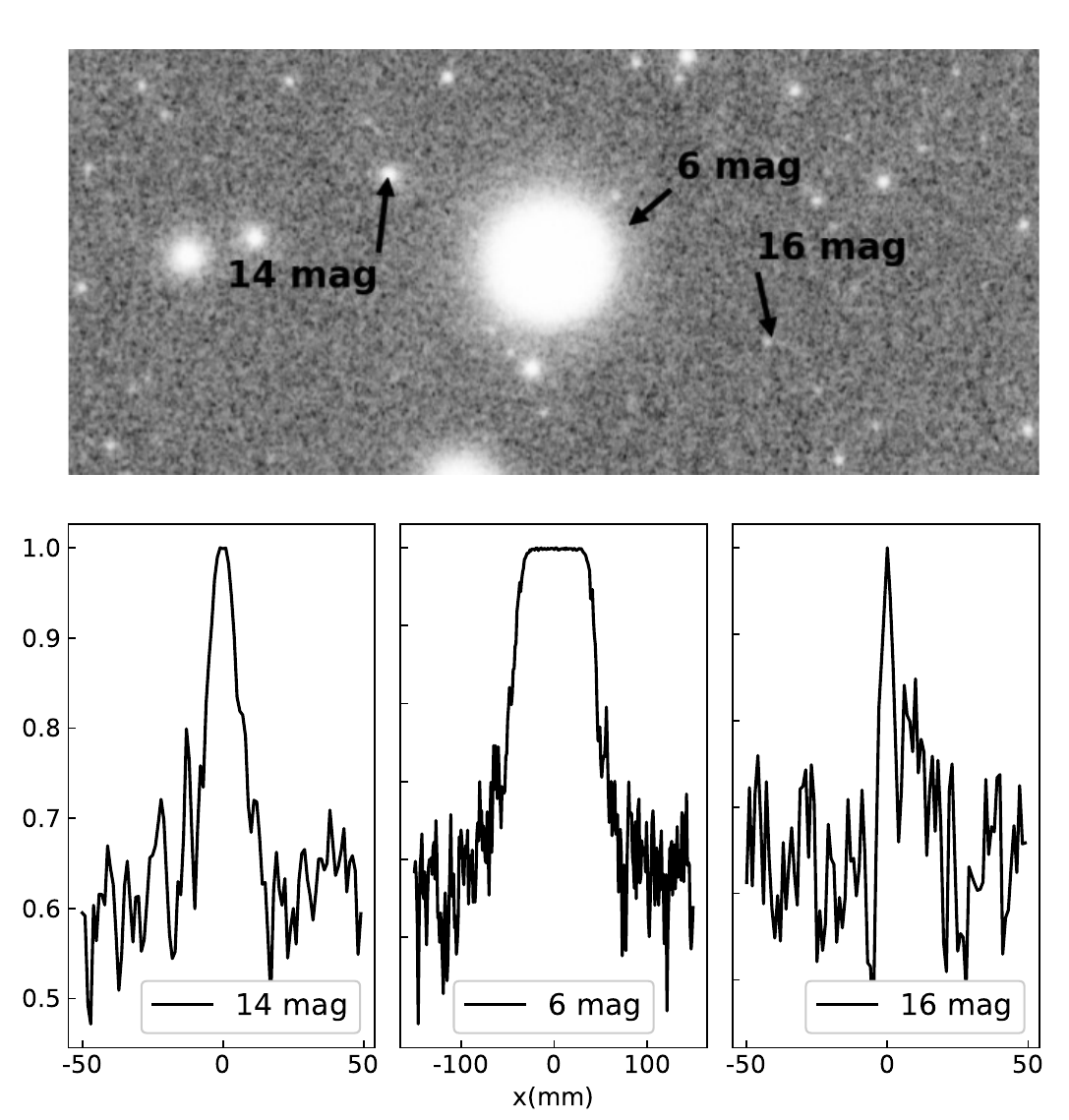}
        \caption{Intensity distributions of stars with different brightness, using plate BJ8304DA3822001 as an example. The upper panel shows the original digitized image of the plate. The lower panel, form left to right, shows the intensity distribution along a horizontal line passing through the center for 14, 6, 16 magnitude objects, respectively. Each distribution has been normalized.}
        \label{Fig14}
    \end{figure}

Figure~\ref{Fig13} shows the distribution of the standard deviation of the positional residuals of the reference stars by telescope with magnitudes. The magnitudes used here are also the $g$ magnitudes of the reference stars in the Gaia DR2 catalog. It can be seen from the figure that the distribution is different for different telescopes. However, there is a general trend that the position residuals of the reference stars first decrease with magnitude and then increase with magnitude, and objects with medium magnitude have the highest position accuracy. The large positional residuals of bright stars may come from inaccurate centering due to saturation, while the large positional residuals of weak stars may come from inaccurate centering due to excessive noise. Medium-brightness stars are less affected by saturation and noise, so they are most accurately centered, and thus have less positional residuals than bright and weak stars. Figure~\ref{Fig14} shows the intensity distributions of stars with different brightness.

Photometric calibration  of photographic glass plate is a complex task in itself, and this is compounded by the missing filter information of the plates. Here we have only roughly fitted the instrumental magnitudes to the $g$ magnitudes of the reference stars in the Gaia DR2 catalog with the $rlowess$ algorithm(\cite{Cleveland+1979}). Detailed photometric calibration is still in progress and will be presented in a subsequent study.

\section{Database}
In this study, we release the digtialized images of all 15,696 single-exposure plates and their corresponding stellar catalogs. The data release of this batch plates, named as China Astronomical Plates Data Release, are archived at the National Astronomical Data Center (NADC) with doi 10.12149/100742 and CSTR 11379.11.100742. Users can access the data through the website https://nadc.china-vo.org/res/r100742/, where the interactive search and download functions on the plates and stellar catalogs are provided.

\section{Conclusion}
\label{sect:conclusion}

To preserve and explore valuable astronomical information on photographic plates, SHAO has organized the transportation of astronomical photographic plates taken at night from various observatories nationwide to the Sheshan Plate Archive for unified preservation. In addition, efforts have been made to compile and analyze plate information. Currently, the Sheshan Plate Archive houses astronomical plates from 11 telescopes, including those from SHAO, NAOC, PMO, YNAO, and QDO, totaling 30,750 plates. These plates span nearly 100 years of astronomical observations.

Based on factors such as degree of damage, mold affected areas, and plate development process, a quality grading system was implemented for the plates. Building upon this grading, cleaning and digital scanning of the plates were carried out, resulting in a final collection of 29,314 digital plate images. Currently, astrometric calibration has been performed on 15,696 single-exposure plates, which involved processes such as object extraction, star matching with the Gaia DR2 catalog, and plate parameter model computation. The results indicate that for long-focus telescopes, such as the 40cm double-tube refracting telescope and the 1.56m reflecting telescope at SHAO and the 1m reflecting telescope at YNAO, the astrometric accuracy of the plates ranges from approximately 0.1\arcsec to 0.3\arcsec. For other medium- and short-focus telescopes, the astrometric accuracy of the plates ranges from 0.3\arcsec to 1.0\arcsec.

In the future, we will try to solve the astrometric calibration of the left 2,580 single-exposure plates, and find the optimal method of photometric calibration for all plates. For the multi-expousred plates, plates capturing near-Earth moving objects and grating observation plates, we will proceed to carry out their astrometric and photometric calibration.

Data relating to this batch of 15,696 single-exposure plates, including digitized images of the plate with observation information and WCS parameters, as well as the scanned images of plate envelopes, are available on the website of the NADC (https://nadc.china-vo.org/res/r100742/). Users can query, retrieve, and download plate data based on keywords such as observatory, telescope, observation year, and observation sky coordinates.

\begin{acknowledgements}

The authors express their sincere gratitude to the NAOC, PMO, YNAO, and QDO for their strong support and assistance in the preservation and transportation of plates. We also thank Professor Songzhu Lan, Professor Guohong Fu, Professor Baoan Yao, Professor Shuhe Wang, Professor Yaqing Mao, Professor Haibin Zhao, Professor Jiexing Yang and Professor Jianwei Zhang for their diligent work organizing the plate information. The authors thank Professor Junliang Zhao, Professor Bochen Qian, and Professor Jiaji Wang for their guidance on plate quality grading. The authors appreciate the advice provided by Professor Wenjing Jin, Professor Yongheng Zhao, Professor Rongchuan Qiao, Professor Zi Zhu and Professor Zhenyu Wu on plate image processing. The authors thank Youfen Wang, Yihan Tao, Linying Mi and Jun Han, for their work on the database. Additionally, we express our sincere gratitude to Professor Hugh Jones for his detailed advice on the content of the article. This work is supported by the Shanghai Science and Technology Innovation Action Plan (Grant No. 21511104100),  the Global Common Challenge Special Project (Grant No. 018GJHZ2023110GC) and the China National Key Basic Research Program (Grant No. 2012FY120500).
\end{acknowledgements}

\bibliographystyle{raa}  
\bibliography{mybib}

\end{CJK*}
\end{document}